# Widely tunable optical parametric oscillation and visible light generation in 4H-SiC microresonators

*Yongsheng Wang, Shuangyou Zhang, Yurong Ren, Paolo Leonelli, Mingjun Chi, and Haiyan Ou**

Department of Electrical and Photonics Engineering, Technical University of Denmark, DK-2800 KGS. Lyngby, Denmark
Email: haou@dtu.dk

Widely separated optical parametric oscillation (OPO) represents a powerful method for coherent wavelength conversion across infrared and visible spectra. While such generation has been demonstrated in material platforms like silicon nitride and lithium niobate, 4H-SiC remains unexplored despite offering combined strong second-order and third-order nonlinearities with ultralow material loss. Here we demonstrate tunable, widely separated OPO generation in 4H-SiC microresonators through dispersion engineering. By optimizing the resonator geometry to achieve normal dispersion at telecommunication wavelengths and pumping at around 1550 nm, a pair of signal and idler spanning nearly an octave is generated, which represents the first demonstration of widely separated OPO in 4H-SiC. The frequency separation is tuned by varying the pump wavelength, with measured signal and idler wavelengths align well with phase-matching prediction. Leveraging the non-centrosymmetric crystal structure of 4H-SiC, the generated OPO signal undergoes cascaded second-harmonic generation (SHG) and sum-frequency generation (SFG) with the pump, yielding coherent visible light at wavelengths below 700 nm. This cascaded upconversion of widely separated OPO signals represents a novel pathway for visible light generation. These results establish 4H-SiC as a promising platform for nonlinear wavelength conversion spanning from visible to 2 μm region.

## 1 Introduction

Optical parametric oscillation (OPO) based on third-order nonlinearity ($\chi^{(3)}$) is an efficient route to on chip wavelength conversion [1, 2, 3, 4, 5]. When the dispersion of a nonlinear resonator is engineered appropriately, widely separated signal and idler frequencies can be generated from a single pump, opening spectral regions that are otherwise difficult to access and enabling photon-pair generation [6]. This approach has been demonstrated across multiple photonic platforms, including silicon nitride [1, 2, 4, 7, 8], magnesium fluoride [9], chalcogenide [5], tantalum pentoxide [10, 11], silicon [12], and aluminum nitride [13], each offering distinct balances of nonlinearity and dispersion control.

On-chip visible-light generation is increasingly important for quantum photonics, sensing, spectroscopy, and metrology [14, 15]. The visible spectral range hosts numerous atomic and molecular transitions that underpin precision spectroscopy, frequency referencing, and optical clock technologies. To access the visible range, different nonlinear processes have been explored across material platforms. Second-harmonic generation (SHG) and sum-frequency generation (SFG), both based on second-order nonlinearity ($\chi^{(2)}$), can upconvert infrared pumps while generally requiring strong $\chi^{(2)}$ and phase-matching conditions [16]. Kerr OPO and stimulated four-wave mixing (FWM), based on third-order nonlinearity ($\chi^{(3)}$), could enable tunable visible-band generation [2, 15]. A particularly attractive route is to combine $\chi^{(3)}$ and $\chi^{(2)}$ in cascaded processes (OPO, SHG and SFG), allowing telecom pumps to be converted into the visible range while extending conversion over an octave. This hybrid strategy leverages the complementary strengths of multiple nonlinear mechanisms and is therefore promising for compact integrated photonics [14, 17, 18].

4H-SiC is a promising platform for this hybrid approach because it offers both significant $\chi^{(2)}$ and $\chi^{(3)}$ nonlinearities together with low material absorption [19, 20, 21]. While OPO around the pump, frequency comb generation, and stimulated Raman lasing have been demonstrated in 4H-SiC [21, 22, 23, 24], widely separated OPO has not yet been reported. Its non-centrosymmetric crystal structure further enables SHG and SFG, making 4H-SiC well suited for combined $\chi^{(2)}$-$\chi^{(3)}$ wavelength conversion [3].

In this work, we demonstrate tunable, widely separated OPO generation in 4H-SiC microring resonators by engineering the geometry to achieve normal dispersion at telecommunication wavelengths [25]. By varying the pump wavelength, we achieve a broad wavelength tuning range spanning from 1200 nm to 2200 nm. The OPO signal then combines with the pump through cascaded SHG and SFG to produce visible light, representing the first demonstration of OPO-assisted visible light generation with telecom pumping in 4H-SiC microring resonators.

## 2 Results

### 2.1 Device Design

Achieving widely separated OPO generation requires careful control of the dispersion of the resonator. Normal dispersion at the pump wavelength is essential to suppress OPO generation near the pump frequency and promote phase-matching for widely separated signal and idler wavelengths [3, 25]. While alternative approaches such as hybrid mode family [12, 26] and photonic crystal-based mode splitting [10, 25] have been demonstrated, engineering normal dispersion through geometric optimization offers a more straightforward path for both design and fabrication. For a microring resonator as presented in Figure 1(a) and (b), geometric parameters including ring width (RW), ring radius (RR), etching depth, and slab thickness collectively determine the dispersion characteristics [2, 10]. Among these dimensions, the radius and ring width have the strongest influence on the $TE_{00}$ mode dispersion. Figure 1(c) presents the simulated dispersion $\beta_2$ at different wavelengths with fixed RW of 2.5 µm but different RRs. Conversely, Figure 1(d) presents the case with varied RWs when the RR is fixed at 60 µm. The etched depth and slab thickness have fixed values of 580 nm and 450 nm respectively. In simulation, the $TE_{00}$ mode can be tuned into the normal dispersion region at telecom wavelength by either increasing the RW or decreasing the RR.

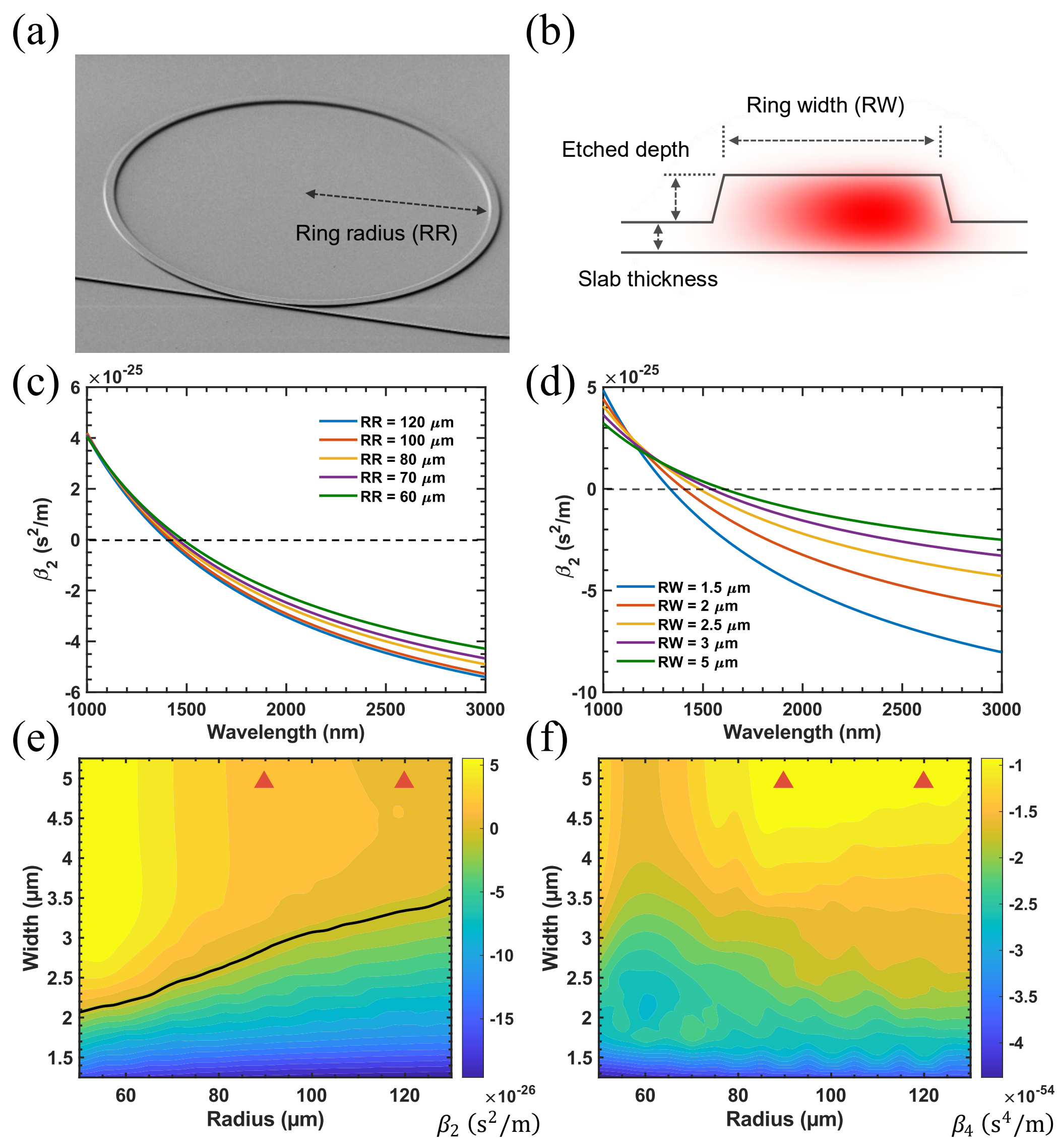

Figure 1: Device design and simulation. a) scanning electron microscope (SEM) image of a fabricated SiC microring resonator. b) simulated $TE_{00}$ mode profile in the ring. c) simulated $\beta_2$ versus wavelengths with fixed RW of 2.5 µm and different RRs at 1550 nm. d) simulated $\beta_2$ versus wavelengths with fixed RR of 60 µm and different RWs at 1550 nm. e) and f) are the simulated $\beta_2$ and $\beta_4$ respectively at 1550 nm in a map of combination of RR from 50 µm to 130 µm and RW from 1.3 µm to 5.2 µm. The solid black line in (e) indicates the zero-dispersion parameters. The two red triangles in (e) and (f) mark the parameters of the ring for single pair OPO and dual pair OPO generation respectively.

The frequency separation in OPO generation is governed by the phase-matching condition, which depends critically on the ratio of the second-order dispersion coefficient $\beta_2$ to the fourth-order coefficient $\beta_4$ [5, 9, 27, 28]. The phase-matching condition for OPO generation in the same mode family could be written as [9, 28]:

$$D_e(\Omega_{\mathrm{pm}})L + 2\gamma PL - \delta_0 = 0, \tag{1}$$

where $\Omega_{\mathrm{pm}}$ is phase-matched frequency, $D_e(\Omega) = \sum_{n=1}^{\infty}(\beta_{2n}\Omega^{2n})/(2n)!$ represents terms with even orders of dispersion, $L$ is the resonator length, $\gamma$ is the nonlinear coefficient, $P$ is the intracavity power and $\delta_0$ is the phase detuning between pump field and cavity resonance. By considering the first two terms of even order dispersion, generating widely separated OPO requires normal dispersion ($\beta_2>0$) and negative fourth-order dispersion ($\beta_4<0$) at pump wavelength. The expected frequency separation

$$\Omega \approx \sqrt{-12\beta_2/\beta_4} \tag{2}$$

indicates that a larger contrast between $\beta_2$ and $\beta_4$ would result in wider frequency separation in OPO. The simulation results presented in Figure 1(e) and (f) map the $\beta_2$ and $\beta_4$ dispersion terms at 1550 nm across RR ranging from 50 µm to 130 µm and RW from 1.3 µm to 5.2 µm. In Figure 1(e), the black line marks the zero-dispersion ($\beta_2 = 0$). The normal dispersion region is the area above the zero-dispersion line. In Figure 1(f), $\beta_4$ within the whole area is negative. A least absolute value of $\beta_4$ could be found in the area with RR larger than 70 µm and RW larger than 4 µm. Therefore, to obtain a widely separated OPO, two devices with different combinations of the parameters are fabricated and tested as the two triangles indicated in Figure 1(e) and (f).

## 2.2 Single Pair OPO Generation

Through a similar fabrication process in [24], we fabricate a microring resonator with a ring width of 5 µm and radius of 120 µm (right triangle in Figure 1(e) and (f)). The etched depth and slab thickness are kept close to the values in simulation. The simulated dispersion indicates the normal dispersion at around 1550 nm of the device in Figure 1(e). The frequency mismatch ($\Delta v = v_s + v_i - 2v_p$) curve predicts OPO signal ($v_s$) and idler ($v_i$) generation at approximately 1150 nm and 2400 nm respectively, when pumped at 1550 nm ($v_p$) as shown in Figure 2(b). In the experimental demonstration presented in Figure 2(a), we pumped the microring at 1567 nm with ~40 mW on-chip power and observed OPO generation with signal and idler wavelengths at approximately 1265 nm and 2059 nm, respectively. The spectra are collected with an optical spectrum analyzer (OSA) covering a range from 1200 nm to 2400 nm. The measured cavity quality factor is approximately 5 million (see Supporting Information) at ~ 1550 nm, resulting in a sufficiently low OPO threshold power [1, 5], thus OPO occurs without requiring an external amplifier at some pump wavelengths, such as 1567 nm.

Tuning the pump wavelength from 1567 nm to 1584 nm produces a continuous decrease in OPO frequency separation, as shown in Figure 2(a). The tunability of widely separated OPO has been reported across different materials [9, 29]. Some methods, such as energy dissipation control, operate in anomalous dispersion regime and exhibit an increasing frequency separation with increasing pump wavelength [5]. In contrast, most reported single-mode-family OPO experiments in normal dispersion regime show a decreasing frequency separation with increasing wavelength, consistent with the behavior observed in the present device [2, 9, 28, 29]. Considering equation (2), the wavelength-dependent frequency separation comes from the wavelength-dependent $\beta_2$. When pumping the device at normal dispersion region, taking the green curve in Figure 1(d) for example, longer wavelength would result in decreasing absolute value of $\beta_2$ (neglecting wavelength-dependent $\beta_4$) and therefore narrower separation between signal and idler.

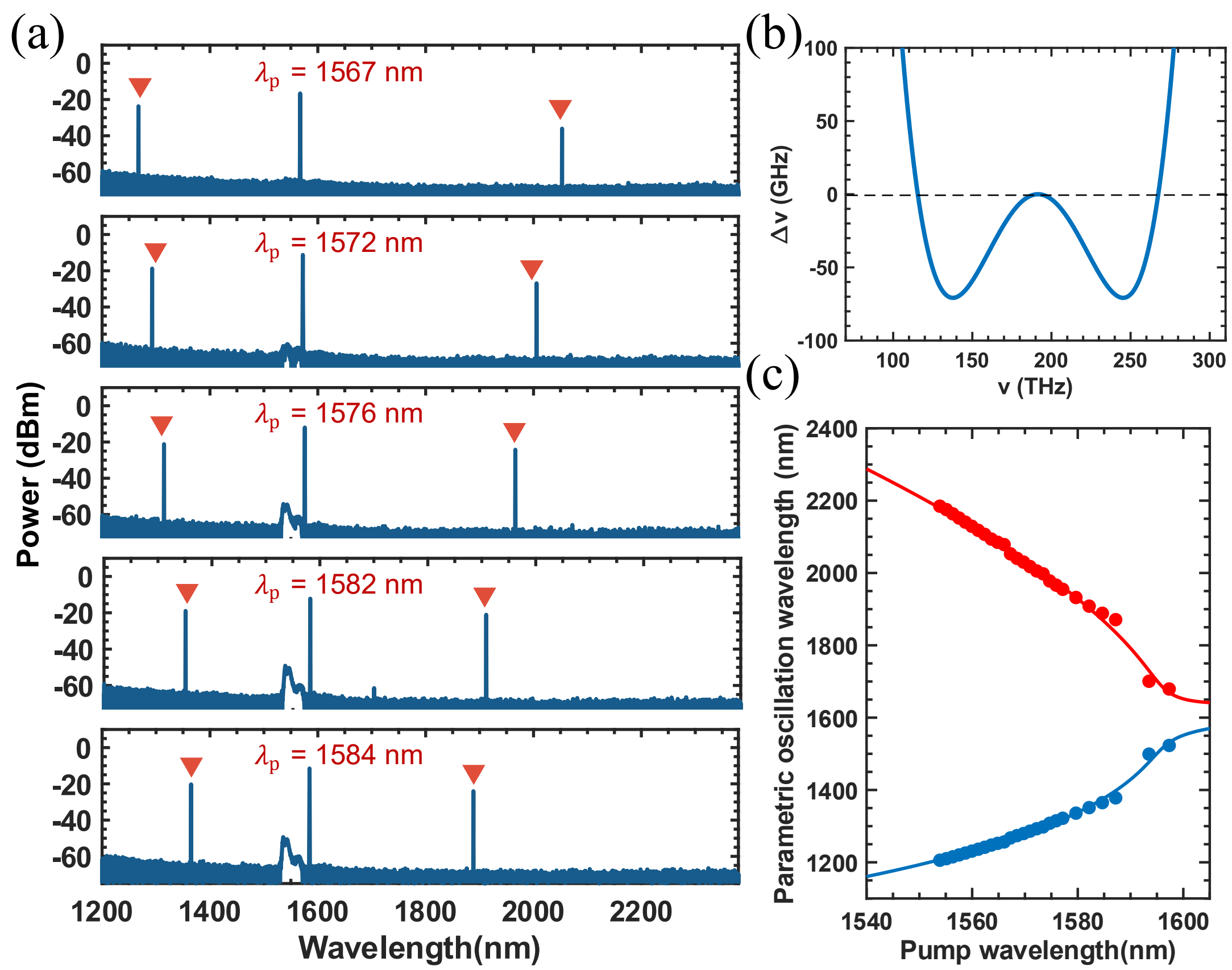

Figure 2: Single pair OPO generation. a) Spectra collected at different pump wavelengths. The signals (shorter wavelength) and idlers (longer wavelength) are marked with red triangles. $\lambda_p$ is the pump wavelength. b) Simulated frequency mismatch curve of $TE_{00}$ with RR of 120 μm and RW of 5 μm when pump the device at 1550 nm. The curve indicates normal dispersion property of the device at 1550 nm with an expected single pair OPO generation. c) Collected parametric oscillation wavelengths versus pump wavelengths. The signal (blue dots) and idler (red dots) fulfill the phase-matching condition at different wavelengths and shows tunability covering from 1200 nm to 2200 nm. The solid curves are fittings by using the phase-matching condition.

Figure 2(c) presents the measured signal (blue dots) and idler (red dots) wavelengths across pump wavelengths ranging from approximately 1550 nm to 1600 nm. With the first two terms of even order dispersion in equation (1), the phase-matching condition becomes [9, 28]

$$\frac{\beta_2 \Omega^2 L}{2} + \frac{\beta_4 \Omega^4 L}{24} + 2\gamma P L - \delta_0 = 0 \tag{3}$$

The solid curves represent the fitting result obtained using the phase-matching condition [9, 30, 31], which shows excellent consistency with the experimental data. By tuning the pump wavelength from 1553 nm to 1587 nm, the signal shift from 1205 nm to 1378 nm correspondingly. The slope of signal branch is about 5.06, which means every 1 nm shift of pump wavelength would lead to 5 nm shift of generated signal. The tuning slope of idler branch is about 9.21, significantly extending the wavelength conversion range.

### 2.3 Dual Pair OPO Generation

Besides the single pair of signal and idler configuration achieved with the 120 μm radius device, we observed 2 pairs of signal/idler generation in a microring resonator with a RR of 90 μm and RW of 5 μm (left triangle in Fig. 1 (e) and (f)). Compared with the single pair case, the dual pair OPO provides additional spectral degrees of freedom, enabling simultaneous access to multiple phase-matched signal and idler channels. The frequency mismatch curve in Figure 3(b) reveals that this device also operates in the normal dispersion region ($\beta_2 > 0$), like the 120 μm radius device. However, the frequency mismatch curve exhibits two zero-crossing points on either side of the pump wavelength. This behavior arises from the phase-matching condition involving the sixth-order dispersion ($\beta_6$) term [1]. For the phase-matching condition (1), the equation now becomes

$$\frac{\beta_2\Omega^2L}{2}+\frac{\beta_4\Omega^4L}{24}+\frac{\beta_6\Omega^6L}{720}+2\gamma PL-\delta_0=0 \quad (4)$$

when $\beta_6$ is considered [9, 28]. The two pairs of zero-crossing points indicate two distinct phase-matching regions on both sides of the pump, where a dual pair widely separated OPO process could be expected.

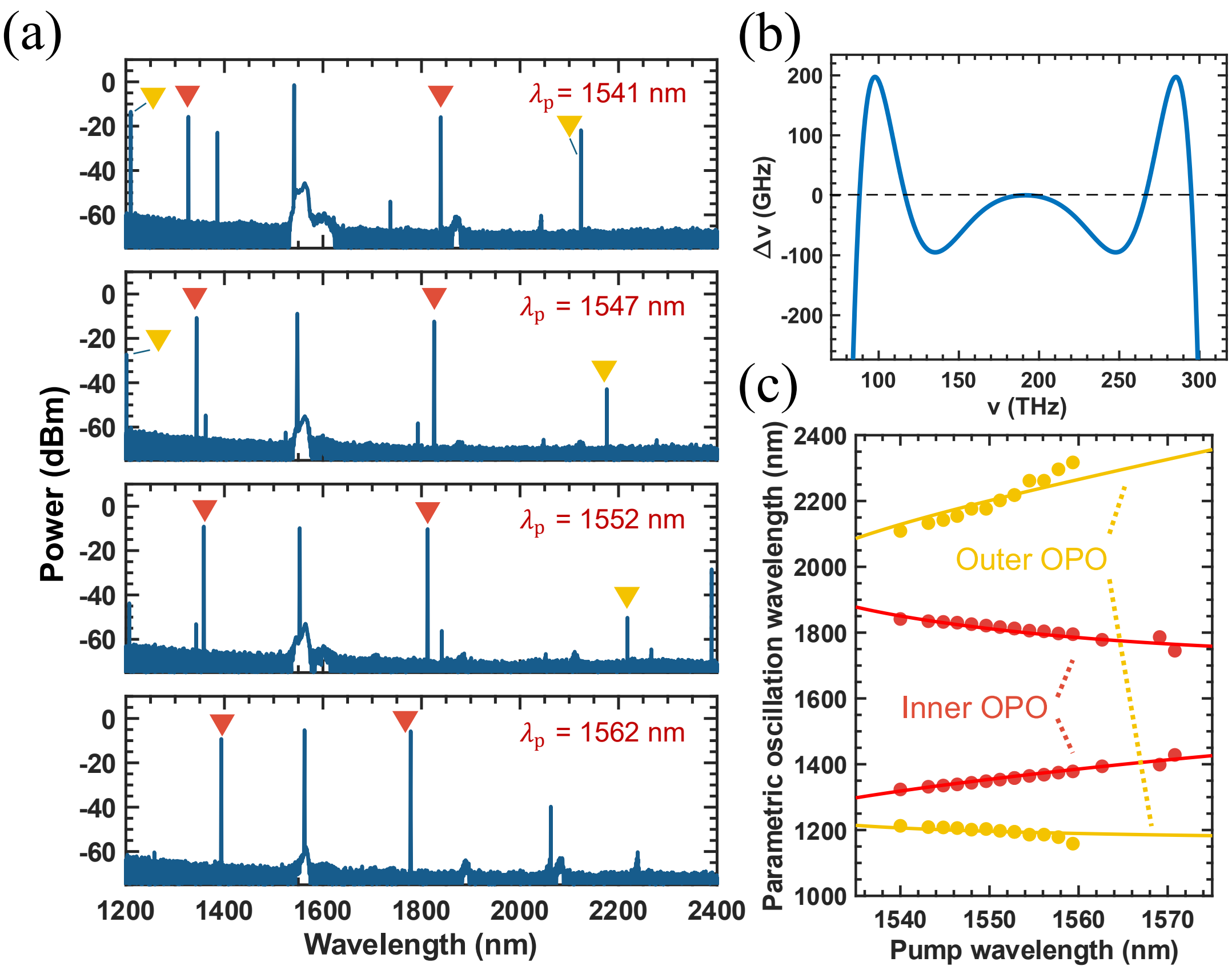

Figure 3: Dual pair OPO generation with RR of 90 μm. a) Spectra collected at different pump wavelengths. When pumping at 1541 nm, two pairs of signal and idler are detected, marked with red (inner OPO) and yellow (outer OPO) triangles. The outer signal (idler) marked with yellow triangles has its wavelength at 1209 nm (2124 nm), fulfills the phase-matching condition with pump wavelength. The inner pair OPO has its signal (idler) at 1327 nm (1839 nm), which also satisfies the phase-matching condition. The weaker peak at 1385 nm is the result of FWM involving the pump, inner signal and outer signal. Other weaker peaks in the spectra come from similar FWM processes. When pumping at 1552 nm, the signal of the outer OPO moves to wavelength shorter than 1200 nm which is out of the OSA range. When pumping at 1562 nm, only the inner OPO is generated. b) Simulated frequency mismatch curve of $TE_{00}$ with RR of 90 μm and RW of 5 μm. Similarly, the curve indicates normal dispersion property of the geometry around 1550 nm but with two expected pairs OPO generation. c) Collected parametric oscillation wavelength versus pump wavelength from 1540 nm to 1570 nm. The solid line is the fitting result to the inner pair (red) and outer pair (yellow) OPO with the phase-matching condition.

In the experimental measurements shown in Figure 3(a), pumping the device at approximately 1541 nm produces inner-sideband OPO (marked with red triangles) and outer-sideband OPO (marked with yellow triangles). This behavior is consistent with previously reported two-pair OPO processes in silicon nitride with strong higher-order dispersion [1]. When pumping the device at 1547 nm, dual pair OPO process still exists while the inner OPO has narrower separation and outer OPO has wider separation compared with the spectrum when pumping at 1541 nm. Specifically, the signal of outer OPO (marked with yellow triangle) has its emission wavelength at 1201 nm, which is approaching the shortest limit wavelength of the OSA. The idler of the outer OPO is located at 2176 nm, correspondingly. Further tuning the pump wavelength to 1552 nm, the frequency separation of the inner OPO (red triangles) keeps decreasing while the outer OPO shows increasing frequency separation. The signal of the outer OPO shifts out of the range of OSA but leaves an artifact signal at ~ 2389 nm (the signal has its actual wavelength at ~1195 nm), which is also observed in previous studies [7]. It is getting more difficult for outer pair OPO generation at longer pump wavelength. The power of outer pair OPO decreases from top to bottom panel in Figure 3(a) and only the inner pair OPO process exists when pumping at 1562 nm. The weaker peaks in the spectra are the results of stimulated FWM between generated signals and pump. For instance, after the generation of two pairs of OPO when pumping at 1541 nm, both signals (1209 nm and 1327 nm) and pump (1541 nm) produce a peak at 1385 nm through FWM process.

Figure 3(c) displays the measured two pairs of the signal and idler wavelength versus pump wavelength from approximately 1540 nm to 1570 nm. The outer pair OPO disappears after 1560 nm due to the shrink gain bandwidth for wider separated OPO process [28, 29] and the higher loss experienced by longer wavelength idlers during oscillation [5]. The solid curves represent fitting result obtained using the phase-matching condition (4) including the 6$^{th}$ order dispersion ($\beta_6$). Following the same fitting method in Figure 2(c), we treat $\beta_2$ as wavelength-dependent while fixing $\beta_4$ and $\beta_6$, which is a simplicity and leads to more derivatives with measured points of outer OPO process as the higher order dispersion would contribute more on wider frequency separation. While the simplified model accurately reproduces the general trend of the experimental data, quantitative agreement is not expected. Deviations may arise from neglected higher-order dispersion terms, fabrication-induced variations in the waveguide geometry, and uncertainties in the material refractive index model across the investigated wavelength range [9].

### 2.4 Visible Light Generation via Cascaded Up Conversion

Besides the widely separated OPO generation, the non-centrosymmetric crystal structure of 4H-SiC enables $\chi^{(2)}$ nonlinear processes that can further extend the utility of these devices. Specifically, the OPO signal and pump photons can further produce light in the visible spectral range through SHG and SFG.

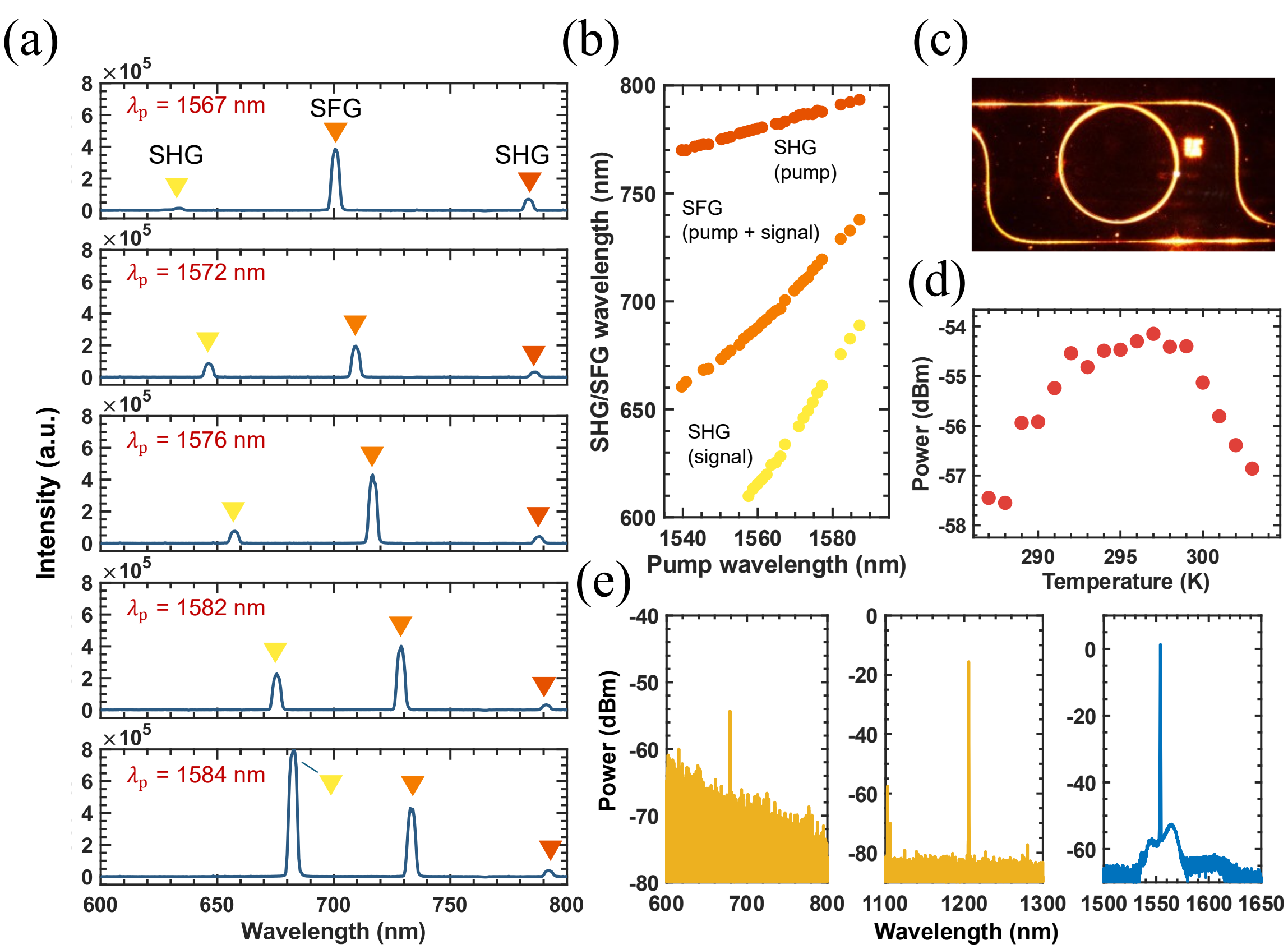

Figure 4: Visible light generation. a) Spectra ranging from 600 nm to 800 nm detected by collecting the scattering light from the same device presented in Figure 2 with a spectrometer. The detected peaks are generated through SHG of the signal (yellow triangles), SFG between signal and pump (orange triangles) and SHG of the pump (red triangles). b) SHG/SFG wavelengths versus pump wavelength. The red points denote the wavelengths indicated by the red triangles in (a), arising from the SHG of the pump. The orange and yellow points similarly correspond to the wavelengths marked by the orange and yellow triangles, respectively. c) A top view optical image of the microring with visible scattering light detected when pumping at 1554 nm. The image was recorded using a color CMOS camera. d) SFG powers collected with OSA at different temperatures. The optimized temperature is ~ 297 K for the most efficient SFG generation. e) Spectra collected around SFG (left), signal (middle) and pump (right) at optimized temperature.

The visible light generation was characterized by collecting scattered light above the microring using a spectrometer. As shown in Figure 4(a), by pumping the same device in Figure 2(a), distinct SHG and SFG peaks are observed in the visible range. The pump wavelength-dependent behavior of these peaks is the result of the tunability of the OPO process. As the pump wavelength increases, all the SHG and SFG peaks shift toward longer wavelengths due to the simultaneous redshift of both pump and signal frequencies. The SHG signal from the pump alone (red triangles) appears at approximately 770–800 nm, which falls outside the visible range. The other two peaks are the components of SHG of signal (yellow triangles) and SFG between signal and pump (orange triangles). The observed intensity variations come from their different phase and resonance matching conditions [32, 33, 34]. Figure 4(b) presents a detailed wavelength map, plotting the observed SFG and SHG peak wavelengths as a function of pump wavelength across the 1540–1590 nm range. The SHG of the signal exhibits a steeper wavelength dependence than SHG of the pump, which is expected because the signal from the OPO process shifts 5 times faster than the pump wavelength in Figure 2(c). This wavelength mapping enables precise control over the visible output color through simple pump wavelength adjustment. Figure 4(c) shows a microscope image of the microring with orange scattering light by pumping at 1554 nm.
To further enhance the power of the visible light, we tune the chip temperature to modulate the resonator properties. Temperature tuning via a thermoelectric controller (TEC) was performed over the range 285–305 K to exploit the temperature-dependent frequency matching of second-order nonlinear processes in the resonator (Figure 4(d)) [32, 33, 34]. The SFG power exhibits a peak enhancement of over 3 dB at approximately 297 K compared to 288 K, demonstrating that thermal control provides a practical mechanism for optimizing visible light output. Figure 4(e) shows the visible spectrum collected when pumping at approximately 1554 nm (right panel) at optimized temperature. The generated signal of the OPO process is at 1205 nm (middle panel). The dominant SFG peak at approximately 678 nm (left panel) exhibits an estimated on-chip power of -35 dBm considering fiber coupling loss. However, temperature tuning here is accompanied by a concurrent modification of the OPO process, as the temperature-induced shift of the signal can transfer it to a different cavity resonance (see Supporting Information), which becomes to satisfy the phase-matching condition at the new temperature.

## 3 Conclusion

In conclusion, we demonstrate the first realization of widely separated optical parametric oscillation in 4H-SiC microring resonators, achieved through careful engineering of the resonator geometry to establish normal dispersion at telecommunication wavelengths. By optimizing the ring radius and ring width, we access distinct regimes of dispersion behavior with one and two pairs of signal/idler generation respectively. Both types of OPO exhibit tunability spanning from approximately 1200 nm to 2200 nm when pumped at around 1550 nm. Besides the widely separated OPO, we also observe the cascaded upconversion to the visible spectral range through simultaneous SHG and SFG processes. These cascaded nonlinear effects enable tunable visible light generation in the orange-to-red range by simply varying the pump wavelength, without requiring tunable sources at shorter wavelengths. Thanks to the strong second- and third-order nonlinearities of 4H-SiC, the ability to achieve wavelength conversion from infrared to visible spectral range within a monolithic device represents a substantial advance for integrated photonics. These results position 4H-SiC microring resonators as promising candidates for on-chip light sources in the visible and mid-infrared range, with applications in quantum photonics, sensing, and nonlinear spectroscopy.

**Supporting Information**
See the Supporting Information for OPO threshold measurement and temperature impact on OPO process.

**Acknowledgements**
Sincere acknowledgments to the Danmarks Frie Forskningsfond (3105-00147B); Horizon 2020 Framework Programme (899679); Villum Fonden (VIL50293); Horizon Europe (101227010).

## References

[1] X. Lu, G. Moille, A. Singh, Q. Li, D. A. Westly, A. Rao, S.-P. Yu, T. C. Briles, S. B. Papp, K. Srinivasan, *Optica* **2019**, *6*, 12 1535.

[2] X. Lu, G. Moille, A. Rao, D. A. Westly, K. Srinivasan, *Optica* **2020**, *7*, 10 1417.

[3] X. Lu, R. M. Gray, J. Stone, S. Zhou, N. Englebert, A. Marandi, K. Srinivasan, *Optica* **2026**, *13*, 1 11.

[4] Y. Sun, J. Stone, X. Lu, F. Zhou, J. Song, Z. Shi, K. Srinivasan, *Light: Science & Applications* **2024**, *13*, 1 201.

[5] D. Xia, J. Zhao, H. Cheng, Z. Wang, J. Huang, L. Luo, D. Liu, S. Yang, B. Zhang, Z. Li, *Laser & Photonics Reviews* **2024**, *18*, 10 2301098.

[6] X. Lu, Q. Li, D. A. Westly, G. Moille, A. Singh, V. Anant, K. Srinivasan, *Nature physics* **2019**, *15*, 4 373.

[7] R. R. Domeneguetti, Y. Zhao, X. Ji, M. Martinelli, M. Lipson, A. L. Gaeta, P. Nussenzveig, *Optica* **2021**, *8*, 3 316.

[8] J. R. Stone, X. Lu, G. Moille, K. Srinivasan, *APL photonics* **2022**, *7*, 12.

[9] N. L. B. Sayson, T. Bi, V. Ng, H. Pham, L. S. Trainor, H. G. Schwefel, S. Coen, M. Erkintalo, S. G. Murdoch, *Nature Photonics* **2019**, *13*, 10 701.

[10] G. M. Brodnik, H. Liu, D. R. Carlson, J. A. Black, S. B. Papp, *Optica* **2025**, *12*, 3 337.

[11] J. A. Black, G. Brodnik, H. Liu, S.-P. Yu, D. R. Carlson, J. Zang, T. C. Briles, S. B. Papp, *Optica* **2022**, *9*, 10 1183.

[12] E. F. Perez, G. Moille, X. Lu, J. Stone, F. Zhou, K. Srinivasan, *Nature communications* **2023**, *14*, 1 242.

[13] Y. Tang, Z. Gong, X. Liu, H. X. Tang, *Optics Letters* **2020**, *45*, 5 1124.

[14] S. Sbarra, B. Zabelich, M. Clementi, J. Liu, T. Kippenberg, C.-S. Br`es, In *Laser Resonators, Microresonators, and Beam Control XXVI*, volume 12871. SPIE, **2024** 1287102.

[15] X. Lu, G. Moille, Q. Li, D. A. Westly, A. Singh, A. Rao, S.-P. Yu, T. C. Briles, S. B. Papp, K. Srinivasan, *Nature Photonics* **2019**, *13*, 9 593

[16] F. Ye, X. Sun, H. K. Tsang, *Optics Letters* **2025**, *50*, 12 4034.

[17] S. Miller, K. Luke, Y. Okawachi, J. Cardenas, A. L. Gaeta, M. Lipson, *Optics Express* **2014**, *22*, 22 26517.

[18] H. Jung, R. Stoll, X. Guo, D. Fischer, H. X. Tang, *Optica* **2014**, *1*, 6 396.

[19] H. Ou, X. Shi, Y. Lu, M. Kollmuss, J. Steiner, V. Tabouret, M. Syv¨aja¨rvi, P. Wellmann, D. Chaussende, *Materials* **2023**, *16*, 3 1014.

[20] X. Shi, A. A. Baiju, V. Dhyani, S. Wang, S. S. Mohanraj, V. Leong, J. Sun, B. Deng, J. Zhang, Y. Wang, et al., *Laser & Photonics Reviews* **2025**, *19*, 17 2500104.

[21] M. A. Guidry, K. Y. Yang, D. M. Lukin, A. Markosyan, J. Yang, M. M. Fejer, J. Vuˇckovi´c, *Optica* **2020**, *7*, 9 1139.

[22] A. A. Afridi, Y. Wang, S. Zhang, R. Wang, J. Li, Q. Li, H. Ou, *Photonics Research* **2025**, *14*, 1 35.

[23] J. Li, R. Wang, A. A. Afridi, Y. Lu, X. Shi, W. Sun, H. Ou, Q. Li, *ACS photonics* **2024**, *11*, 2 795.

[24] Y. Wang, S. Zhang, Y. Lu, A. A. Afridi, X. Shi, Y. Ren, M. Chi, K. Rottwitt, H. Ou, *Applied Physics Letters* **2026**, *128*, 18 181104.

[25] X. Lu, A. Chanana, F. Zhou, M. Davanco, K. Srinivasan, *Optics letters* **2022**, *47*, 13 3331.

[26] F. Zhou, X. Lu, A. Rao, J. Stone, G. Moille, E. Perez, D. Westly, K. Srinivasan, *Laser & Photonics Reviews* **2022**, *16*, 7 2100582.

[27] J. D. Harvey, R. Leonhardt, S. Coen, G. K. Wong, J. Knight, W. J. Wadsworth, P. St. J. Russell, *Optics letters* **2003**, *28*, 22 2225.

[28] N. L. B. Sayson, K. E. Webb, S. Coen, M. Erkintalo, S. G. Murdoch, *Optics Letters* **2017**, *42*, 24 5190.

[29] S. Fujii, S. Tanaka, M. Fuchida, H. Amano, Y. Hayama, R. Suzuki, Y. Kakinuma, T. Tanabe, *Optics letters* **2019**, *44*, 12 3146.

[30] J. Li, R. Wang, L. Cai, Q. Li, *Physical Review Applied* **2023**, *19*, 3 034083.

[31] X. Shi, W. Fan, Y. Lu, A. K. Hansen, M. Chi, A. Yi, X. Ou, K. Rottwitt, H. Ou, *APL Photonics* **2021**, *6*, 7.

[32] X. Wu, Z. Hao, L. Zhang, D. Jia, R. Ma, C. Tao, F. Gao, F. Bo, G. Zhang, J. Xu, *Laser & Photonics Reviews* **2024**, *18*, 7 2300951.

[33] B. Wang, Y. Ji, L. Gu, L. Fang, X. Gan, J. Zhao, *ACS Photonics* **2022**, *9*, 5 1671.

[34] J. Lu, J. B. Surya, X. Liu, A. W. Bruch, Z. Gong, Y. Xu, H. X. Tang, *Optica* **2019**, *6*, 12 1455.